\documentclass[aps,eqsecnum,preprint,floats,epsf,epsfig,nofootinbib,11pt]{revtex4}
\usepackage{epsfig,color,amsmath,multirow}

\begin{document}

\title{\bf Doubly-Heavy Baryon Weak Decays:  $\Xi_{bc}^{0}\to pK^{-}$ and $\Xi_{cc}^{+}\to \Sigma_{c}^{++}(2520)K^{-}$  }
\author{Run-Hui Li$^1$~\footnote{Email:lirh@imu.edu.cn}, Cai-Dian L\"u$^2$~\footnote{Email:lucd@ihep.ac.cn}, Wei Wang$^3$~\footnote{Email:wei.wang@sjtu.edu.cn}, Fu-Sheng Yu$^4$~\footnote{Email:yufs@lzu.edu.cn}, Zhi-Tian Zou$^5$~\footnote{Email:zouzt@ytu.edu.cn}}
\affiliation{\small
$^{1}$School of Physical Science and Technology, Inner Mongolia University, Hohhot 010021, China;
\\
$^2$Institute of High Energy Physics, YuQuanLu 19B, Beijing 100049, China; \\
School of Physics, University of Chinese Academy of Sciences, YuQuanLu 19A, Beijing 100049, China; \\
$^{3}$INPAC, Shanghai Key Laboratory for Particle Physics and Cosmology, Department of Physics and Astronomy, Shanghai Jiao-Tong University, Shanghai 200240, China;\\
$^{4}$School of Nuclear Science and Technology, Lanzhou University,
Lanzhou, 730000, China;\\
Research Center for Hadron and CSR Physics, Lanzhou University
and Institute of Modern Physics of CAS, Lanzhou, 730000, China;\\
$^5$Department of Physics, Yantai University, Yantai 264055, China
}

\begin{abstract}
Doubly-heavy baryons, with  two heavy  and  one light quarks,  are   expected to exist in QCD and their masses  have been predicted in the quark model. However their existence is  not well established so far  in experiment.  In this work, we explore the possibility of searching for $\Xi_{bc}$ and $\Xi_{cc}^{+}$ in the  $W$-exchange processes, $\Xi_{bc}^{0}\to pK^{-}$ and $\Xi_{cc}^{+}\to \Sigma_{c}^{++}(2520)K^{-}$. On the basis of perturbative calculations, we estimate  the branching ratio of the first decay  as  ${\cal BR}(\Xi_{bc}^0\to p^+ K^-)\approx3.21\times {\cal R}^2_f \times {\cal R}_{\tau}\times 10^{-7}$, where ${\cal R}_f$ (${\cal R}_{\tau}$) are the ratios of the decay constants (lifetimes) of $\Xi_{bc}^{0}$ and $\Lambda_b^{0}$. The branching ratio of $\Xi_{cc}^{+}\to \Sigma_{c}^{++}(2520)K^{-}$ is related to that  of $\Lambda_c^+\to \Delta^{++} K^-$, and thereby a conjectured topology analysis   leads to the range for the branching ratio as:  ${\cal BR}(\Xi_{cc}^+\to \Sigma_{c}^{++}(2520) K^-)\in \left[0.36\%,1.80\%\right]$. The  decay $\Xi_{cc}^+\to \Sigma_{c}^{++}(2520) K^-$ would be  reconstructed in the $ \Lambda_{c}^{+}K^{-}\pi^{+}$ final state which is easy to access even at a hadron collider. Based on the two facts that abundant  heavy quarks can be produced at a hadron collider like LHC, and   the branching ratios of $\Xi_{bc}^{0}\to pK^{-}$ and $\Xi_{cc}^{+}\to \Sigma_{c}^{++}(2520)K^{-}$ are sizable,  we urge our experimental colleagues to perform a search at LHCb. This will presumably  lead to the discovery of the $\Xi_{bc}$ and $\Xi_{cc}^{+}$ , and precision measurements of the branching ratios in the future are helpful to investigate  their decay mechanism.

\end{abstract}

\maketitle

\section{Introduction}

In  the  quark model, two heavy quarks (bottom and/or charm) can bound together  with a light quark to  form the so-called doubly-heavy  baryons.   The study  of doubly-heavy  baryons is of great interest for the understanding of the hadron spectroscopy of various systems. The study of their decays can also shed light on  the nonperturbative dynamics in the transition with two  heavy quarks.

The mass spectra of ground states of   doubly-heavy baryons have already been studied in various versions of the quark model~\cite{Kiselev:2001fw,Karliner:2014gca,DeRujula:1975qlm,Anikeev:2001rk,Fleck:1989mb,Korner:1994nh,Roncaglia:1995az,Lichtenberg:1995kg,Ebert:1996ec,Gerasyuta:1999pc,Itoh:2000um,Narodetskii:2002ib,Ebert:2002ig,He:2004px,Richard:2005jz,Migura:2006ep,Albertus:2006ya,Roberts:2007ni,Weng:2010rb,Zhang:2008rt,Lewis:2001iz,Flynn:2003vz,Liu:2009jc,Namekawa:2012mp,Alexandrou:2012xk,Briceno:2012wt,Alexandrou:2014sha,Gerasyuta:2008zy,Garcilazo:2016piq,Valcarce:2008dr}; meanwhile their lifetimes \cite{Karliner:2014gca,Anikeev:2001rk,Kiselev:2001fw,Moinester:1995fk,Chang:2007xa,Guberina:1999mx} and production~ \cite{Kiselev:2001fw,Karliner:2014gca,Chang:2006eu,Chang:2006xp,Gunter:2001qy,Kiselev:1994pu,Koshkarev:2016rci,Koshkarev:2016acq,Ma:2003zk,Chang:2009va,Chang:2007pp,Zhang:2011hi,Chen:2014hqa} are also investigated in phenomenological ways.  However the doubly-heavy baryons like $\Xi_{cc, bc}$ and $\Omega_{cc,bc}$  are not well established in experiment. The  only evidence for  $\Xi_{cc}^{+}$,  found by the SELEX collaboration \cite{Mattson:2002vu,Ocherashvili:2004hi}, is not confirmed by other experiments \cite{Kato:2013ynr,Aaij:2013voa}. Actually,  the search for doubly-heavy baryons depends on two factors, the production and decay. At a hadron collider like LHC, abundant heavy quarks can be generated, which means plenty of doubly-heavy hadrons due to  quark-hadron duality. For instance the properties of the $B_{c}^{\pm}$ have recently  been   studied by the LHCb collaboration in great detail~\cite{Aaij:2016qlz,Aaij:2017kea,Aaij:2013cda,Aaij:2016xas,Abazov:2008kv,Aaij:2014ija}. Another factor, the decay  of the doubly-heavy baryons, is the main focus of this paper and subsequent ones.

In this work,  we will  explore the possibility of searching for the  $\Xi_{bc}$ and $\Xi_{cc}$ through the $\Xi_{bc}^{0}\to pK^{-}$ and $\Xi_{cc}^{+}\to \Sigma_{c}^{++}(2520)K^{-}\to \Lambda_{c}^{+}\pi^{+}K^-$ decays.  Both channels are  dominated by the $W$-exchange contribution in theory, but as we will demonstrate later that their  branching ratios are likely sizable.  These channels have an advantage because of the all charged final states and thus can be accessed straightforwardly in experiment.

So far, our knowledge of the doubly-heavy baryon decays is limited; for example even the decay constants are still not well known, which prevents us from making reliable predictions of decay branching ratios in a QCD-rooted approach. Fortunately one can make use of the analogue in $\Lambda_b$ and $\Lambda_c$ decays, which can result in  an estimation of decay branching ratios.  Since the bottom quark is annihilated, there is a large energy release in the decay $\Xi_{bc}^{0}\to pK^{-}$.  As a result, the proton and kaon move very fast in the rest frame of $\Xi_{bc}$. The large momentum transfer in this process guarantees the  applicability of QCD perturbation theory. It enables us to relate this channel to  $\Lambda_{b}^{0}\to p\pi^{-}$ which has been calculated in perturbative QCD~\cite{Lu:2009cm}.
The decay $\Xi_{cc}^{+}\to \Sigma_{c}^{++}(2520)K^{-}$ is governed by the $c\to su\bar d$ transition at the quark level, where the momentum transfer is limited and thus nonperturbative dynamics is dominating. The decay $\Lambda_{c}^{+}\to \Delta^{++}K^{-}$ ~\cite{Olive:2016xmw}, which is also a pure $W$-exchange process with exactly the same polarization contributions, provides an opportunity to estimate the branching ratio of $\Xi_{cc}^{+}\to \Sigma_{c}^{++}(2520)K^{-}$.

In the rest of this paper, we  will analyze the $\Xi_{bc}^{0}\to pK^{-}$ decay in Sec. \ref{sec:bc} and $\Xi_{cc}^{+}\to \Sigma_{c}^{++}(2520)K^{-}$ decay in Sec. \ref{sec:cc}. A brief summary is given in the last section.

\section{The study of $\Xi_{bc}^{0}\to pK^{-}$ decay}
\label{sec:bc}

\begin{figure}[phtb!]
\begin{center}
\includegraphics[scale=0.44]{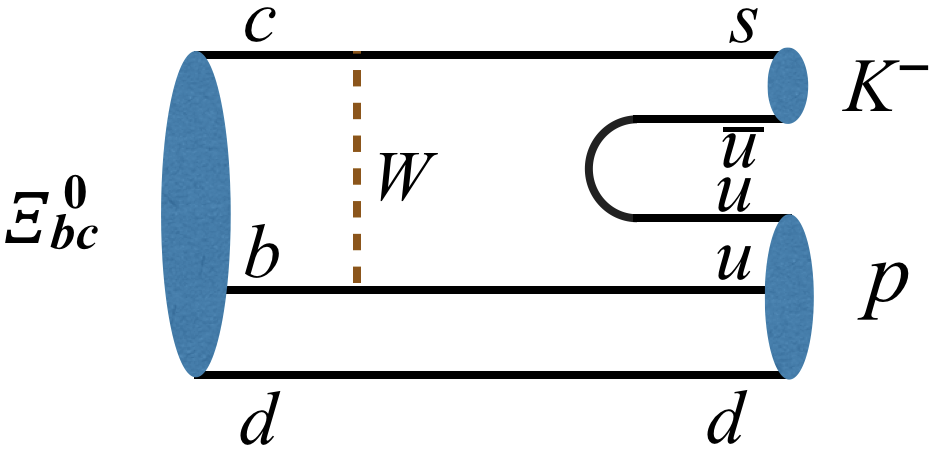}
\hspace{0.5cm}
\includegraphics[scale=0.51]{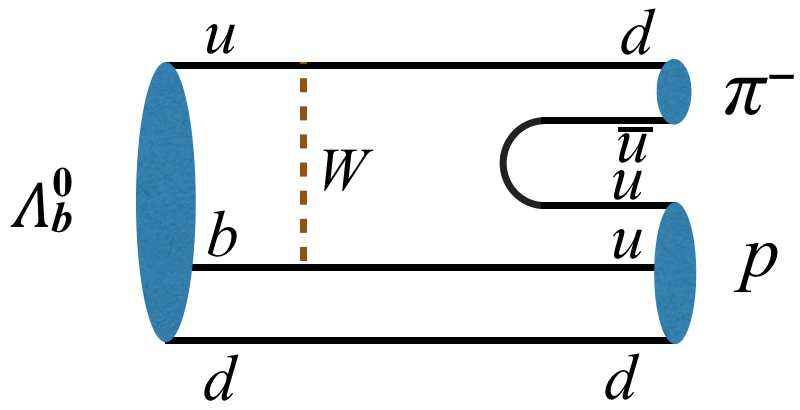}
\hspace{5cm}(a) \hspace{7cm}(b)
\caption{The leading $W$-exchange diagram contributions for $\Xi_{bc}^{0}\to pK^{-}$ and $\Lambda_b\to p \pi^-$ decays.}
\label{fig:wexb}
\end{center}
\end{figure}

The decay $\Xi_{bc}^0\to pK^-$ can proceed either through  a $W$ exchange between $b\to u$ and $c\to s$ transitions or by double flavor changing neutral current (FCNC) processes of $b \to s$ and $c\to u$. The latter one is highly suppressed by loop effects and thus  can be neglected. Therefore the $W$-exchange mechanism depicted in Fig. \ref{fig:wexb}(a) is the only tree level contribution to the decay $\Xi_{bc}^0\to pK^-$. The amplitude of $\Xi_{bc}^0\to pK^-$ decay can be decomposed into two different structures with corresponding coefficients $f_1$ and $f_2$ \cite{Lu:2009cm}:
 \begin{equation}
 {\cal M}=\bar p (p^\prime)[f_1 + f_2 \gamma_5]\Xi_{bc}(p)\, ,\label{eq:amp}
 \end{equation}
where $\bar p (p^\prime)$ and $\Xi_{bc}(p)$ are the respective spinors of the proton and $\Xi_{bc}$ baryon with $p^\prime$ and $p$ being the momenta. Since $\Xi_{bc}$ is very  heavy, $m_{\Xi_{bc}}=6.80\pm 0.05{\rm GeV}$~\cite{Kiselev:1999zj}, the energy release of this decay is expected to be large. Therefore the proton and kaon in the final state can be approximately treated as light-like particles, and in the light cone coordinates the momenta $p$ and $p^\prime$ are defined as
\begin{equation}
p=\frac{m_{\Xi_{bc}}}{\sqrt{2}}(1,1,{\bf 0_T})\,\, , \,\, p^\prime=\frac{m_{\Xi_{bc}}}{\sqrt{2}}(0,1,{\bf 0_T})\,\, . \label{eq:momenta}
\end{equation}
$f_1$ and $f_2$ in Eq. (\ref{eq:amp}) can be extracted from the calculation of Fig. \ref{fig:wexb}(a). Firstly a very simple picture in the heavy quark limit is adopted, in which $b$-quark is the only heavy one while $u$, $d$, $s$ and $c$ quarks are massless. At  the leading power in  $m_c/m_b$, one can find a very similar diagram, depicted  by Fig. \ref{fig:wexb}(b), in the decay $\Lambda_b^0 \to p^+ \pi^-$. Factoring out  the  decay constants, masses and CKM matrix elements, one can find that the two diagrams in Fig. \ref{fig:wexb} should contribute equally.
Fig. \ref{fig:wexb}(b), called  \emph{Bow-tie} in Ref.~\cite{Lu:2009cm} but  named as the \emph{W-exchange} ($E$ for short) contribution in this paper, has been calculated in  the conventional perturbative QCD approach. For the pQCD approach, see Ref.~\cite{Keum:2000ph,Lu:2000em}. For the decay $\Lambda_b^0 \to p^+ \pi^-$, the $E$ diagrams give~\cite{Lu:2009cm}
\begin{equation}
f_1=-7.00\times 10^{-11} + i 3.33\times 10^{-10} \, ,\, f_2=2.21\times 10^{-10} -i 4.04\times 10^{-11}.\label{eq:f1f2w}
\end{equation}
In contrast with the situation in $\Xi_{bc}^0 \to p^+ K^-$ decay, both tree (shown in Fig. \ref{fig:wexb}(b)) and   penguin operators contribute to the $E$ topological diagrams of $\Lambda_b^0 \to p^+ \pi^-$ decay.  One can see from Table I of Ref.~\cite{Lu:2009cm} that the penguin contributions are suppressed by  at least one order of magnitude  due to the Wilson coefficients.   Therefore, one can take the values of $f_1$ and $f_2$ in Eq. (\ref{eq:f1f2w}) approximately as the contribution due to Fig. \ref{fig:wexb}(b). The differences  between diagram (a) and (b) of Fig. \ref{fig:wexb} are listed as follows.
\begin{itemize}
 \item The CKM matrix elements, which is $V_{ub}V_{cs}^*$ in diagram (a) and $V_{ub}V_{ud}^*$ in diagram (b),  are approximately the same in magnitude.
 \item The kaon and pion are in the same octet  in the $SU(3)$ symmetry. In this paper the $SU(3)$ symmetry breaking  effect arises from  the decay constants which  are used as  $f_{\pi}=130\,{\rm MeV},\, f_{K}=156\,{\rm MeV}$~\cite{Olive:2016xmw}.
 \item The difference between $\Xi_{bc}$ and $\Lambda_b$ resides  in the decay constants, masses and lifetimes.
  ${\cal R}_f\equiv f_{\Xi_{bc}}/f_{\Lambda_b}$ and ${\cal R}_{\tau}=\tau_{\Xi_{bc}}/\tau_{\Lambda_b}$ will appear as unknown parameters because of the absence of $\Xi_{bc}$'s decay constant and the large ambiguity of its predicted lifetime in the literature. According to the structure of pseudoscalar meson wave functions, the terms in the magnitude can be divided into two groups: one with the chiral mass of pseudoscalar meson, the other without. The former group is proportional to $m^5_{\Lambda_b/\Xi_{bc}}r_0$ with $r_0=m_{0\pi/K}/m_{\Lambda_b/\Xi_{bc}}$, and the latter one is proportional to $m^5_{\Lambda_b/\Xi_{bc}}$. Here $m_{0\pi/K}=m_{\pi/K}^2/(m_{q_1}+m_{q_2})$ is the chiral mass of the pseudoscalar meson, where $m_{q_1}$ and $m_{q_2}$ are the masses of the valence quarks. Neglecting the small difference between $r_0=m_{0K}/m_{\Xi_{bc}}$ and $r_0=m_{0\pi}/m_{\Lambda_b}$, one can treat the total magnitude as simply being proportional to $m^5_{\Lambda_b/\Xi_{bc}}$.
\end{itemize}
Combining all the pieces, the $f_{1,2}$ for $\Xi_{bc}^0\to p^+K^-$ decay are given as
\begin{equation}
f_1=\left(-2.18\times 10^{-10} + i 1.04\times 10^{-9}\right){\cal R}_f \, ,\, f_2=\left( 6.88\times 10^{-10} - i 1.26\times 10^{-10}\right){\cal R}_f,
\end{equation}
and the branching ratio is predicted as
\begin{equation}
{\cal BR}(\Xi_{bc}^0\to p^+ K^-)\approx3.21\times {\cal R}^2_f \times {\cal R}_{\tau}\times 10^{-7}.
\label{eq:bctopk}
\end{equation}
The predicted $\tau_{\Xi_{bc}}$ ranges from $0.1$ to $0.3\times10^{-12}$s~\cite{Karliner:2014gca,Anikeev:2001rk,Kiselev:2001fw}, and $\tau_{\Lambda_b}=(1.466\pm0.010)\times10^{-12}$s~\cite{Olive:2016xmw}. Assuming the decay constants of $\Xi_{bc}^0$ and $\Lambda_b^0$ are of the same order, it is expected that the branching ratio of $\Xi_{bc}^0\to p^+ K^-$ is of the order of $10^{-7\sim-8}$.

It should be stressed that the perturbative QCD calculation bases on the leading order analysis in the $1/m_b$ expansion.  Above all, the study of $\Lambda_b$ decays shows  that only considering the perturbative contribution~\cite{Lu:2009cm} will undershoot the data, and it indicates that the nonperturbative mechanism might also contribute sizably. If it were also the situation in  $\Xi_{bc}^0\to p^+ K^-$, the  branching ratio  will be greatly enhanced compared to the value given in Eq. (\ref{eq:bctopk}).

\section{The Study of $\Xi_{cc}^{+}\to \Sigma_{c}^{++}(2520)K^{-}$ decay}
\label{sec:cc}

\begin{figure}[phtb!]
\begin{center}
\includegraphics[scale=0.5]{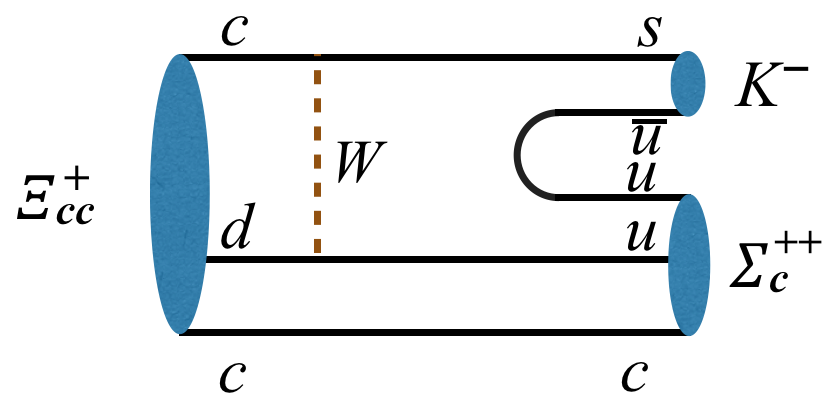}
\includegraphics[scale=0.5]{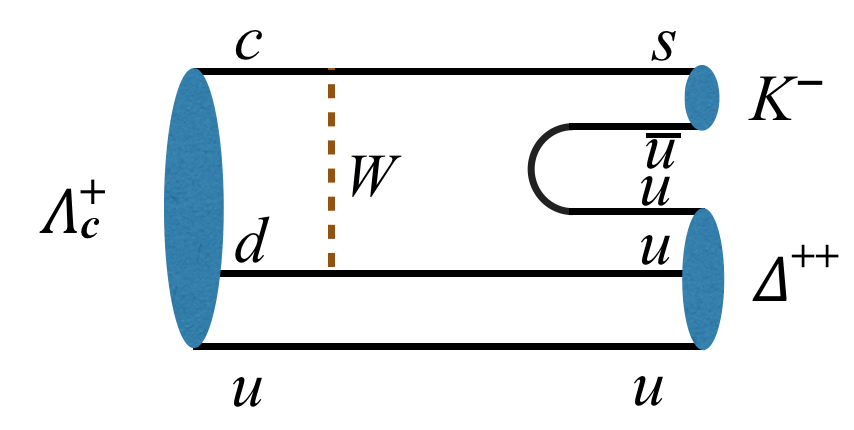}
\hspace{5cm}(a)\hspace{7cm}(b)
\caption{Leading Feynman diagrams of $\Xi_{cc}^{+}\to \Sigma_{c}^{++}(2520)K^{-}$ and $\Lambda_{c}^{+}\to \Delta^{++}K^{-}$ decays.}\label{fig:cc}
\end{center}
\end{figure}

Another decay with only the $E$ contribution at tree level is $\Xi_{cc}^{+}\to \Sigma_{c}^{++}(2520)K^{-}$ (shown in Fig. \ref{fig:cc}(a)). The $\Xi_{cc}$ with two identical heavy quarks is expected to be quite different from $\Xi_{bc}$. Nonperturbative contributions are much more important because of the low energy scale. Fortunately, there exists a twin process, $\Lambda_c^{+}\to \Delta^{++}K^{-}$, which is depicted in Fig. \ref{fig:cc}(b). Both of the processes are decays of a $\frac{1}{2}^+$ baryon to a $\frac{3}{2}^+$ baryon and a kaon. The only difference is the \emph{spectator} quark. Investigating these two decays in the rest frame of the initial baryons, one finds that either $\Sigma_c^{++}(2520)$ or $\Delta^{++}$ in the final states moves slowly, since the energy release is very small. It means that the spectator $c$-quark almost keeps static in the decay $\Xi_{cc}^{+}\to \Sigma_{c}^{++}(2520)K^{-}$. Meanwhile, the spectator $u$-quark moves with the similar energy of $\Lambda_{\rm{QCD}}$ in both initial and final states. Both the spectators exchange little energy with the weak transitions, which indicates that the magnitudes of the diagrams are not sensitive to the spectators. With this viewpoint, one can reach the conclusion that Fig. \ref{fig:cc}(a) and (b) have nearly equal magnitudes. However, if comparing their branching ratios, one should notice that each charm quark in $\Xi_{cc}$ can decay to a strange one. Combing the differences of phase space and the lifetimes, the branching ratio is given by
\begin{align}
{\cal BR}(\Xi_{cc}^+\to \Sigma_{c}^{++}(2520) K^-)&\approx{\cal BR}(\Lambda_c^+\to \Delta^{++} K^-)\times 0.66\times 4\times\frac{1}{2}\times{\cal R}_{\tau}^\prime
\nonumber\\
&={\cal BR}(\Lambda_c^+\to \Delta^{++} K^-)\times 1.32\times{\cal R}_{\tau}^\prime,
\end{align}
where $0.66$ is from the phase space difference, $\frac{1}{2}$ from the symmetry of exchanging identical particles $c$-quarks in the $\Xi_{cc}$, and ${\cal R}_{\tau}^\prime=\tau_{\Xi_{cc}}/\tau_{\Lambda_c}$ is the ratio of $\Xi_{cc}$ and $\Lambda_c$ lifetimes. There are a lot of studies of the mass of $\Xi_{cc}$ with small deviations from each other, and $m_{\Xi_{cc}}=3.627\pm0.012\,\rm{GeV}$~\cite{Karliner:2014gca} is adopted in this paper.
The masses which are not listed here are all from the Particle Data Group~\cite{Olive:2016xmw}.
Using the measured value of branching ratio
${\cal BR}(\Lambda_{c}^{+}\to \Delta^{++}K^{-})=(1.09\pm0.25)\%$~\cite{Olive:2016xmw},
we have
\begin{align}
{\cal BR}(\Xi_{cc}^+\to \Sigma_{c}^{++}(2520) K^-)=(1.44\pm0.33)\%\times{\cal R}_{\tau}^\prime.
\end{align}
$\tau_{\Xi_{cc}}$ is predicted in the literature with large uncertainty, ranging from 0.5 to 2.5$\times10^{-13}\,\rm{s}$~ \cite{Karliner:2014gca,Anikeev:2001rk,Kiselev:2001fw,Moinester:1995fk,Chang:2007xa,Guberina:1999mx}, and $\tau_{\Lambda_c}=(2.00\pm0.06)\times 10^{-13}\,\rm{s}$~\cite{Olive:2016xmw}. With the experimental errors neglected, the branching ratio is given by
\begin{equation}
{\cal BR}(\Xi_{cc}^+\to \Sigma_{c}^{++}(2520) K^-)\in \left[0.36\%,1.80\%\right]. \label{brs}
\end{equation}

 Experimentally, $\Sigma_c^{++}(2520)$ is found to decay into $\Lambda_c^+\pi^+$ with branching ratio $100\%$. This  would make  the branching ratio of the three-body decay  $\Xi_{cc}^+\to \Lambda_c^+K^-\pi^+$ the same order as  that of the two body decay $\Xi_{cc}^+\to \Sigma_{c}^{++}(2520) K^-$  shown in Eq.~(\ref{brs}). Therefore, this decay mode is  an ideal discovery channel. Actually, the SELEX collaboration found the evidence for $\Xi_{cc}^+$ in this process~\cite{Mattson:2002vu}.

\section{Summary}

The  existence of doubly-heavy baryons which consist of two heavy quarks and a light one is undoubted in QCD but has never been confirmed in experiment. Searching for these baryons is of great interest in hadron physics, and we believe that it is only a problem of time to establish their existence. To improve the efficiency of experimental searching, it becomes urgent to study theoretically the doubly-heavy baryon decays. In this work we present an estimate of the branching ratio of $\Xi_{bc}^{0}\to pK^{-}$ and $\Xi_{cc}^{+}\to \Sigma_{c}^{++}K^{-}$ decays, which may be ideal channels for the $\Xi_{bc}$ and $\Xi_{cc}$ reconstruction.

On basis of perturbative calculation, the first branching ratio is given by ${\cal BR}(\Xi_{bc}^0\to p^+ K^-)\approx3.21\times {\cal R}^2_f \times {\cal R}_{\tau}\times 10^{-7}$, where ${\cal R}_f$ (${\cal R}_{\tau}$) are the ratios of the decay constants (lifetimes) of $\Xi_{bc}$ and $\Lambda_b$. Considering that the lifetime of $\Xi_{bc}^0$ is smaller than that of $\Lambda_b$ by  one order of magnitude, and assuming their decay constants are at the same order, the branching ratio of $\Xi_{bc}^0\to p^+ K^-$ is of order of $10^{-7}\sim 10^{-8}$. The analysis of $\Lambda_b$ decays indicates that the decay magnitudes are probably underestimated if only considering the perturbative contribution. Including  the nonperturbative contribution,  one may get a larger result for the branching ratio.

$\Xi_{cc}^{+}\to \Sigma_{c}^{++}(2520)K^{-}$ is dominated by the nonperturbative dynamics because of small energy release. By analyzing an analogous decay $\Lambda_c^+\to \Delta^{++} K^-$, and utilizing its experimental result, we estimate that  ${\cal BR}(\Xi_{cc}^+\to \Sigma_{c}^{++}(2520) K^-)=(1.44\pm0.33)\%\times{\cal R}_{\tau}^\prime$, where ${\cal R}_{\tau}^\prime$ is the ratio of $\Xi_{cc}$ and $\Lambda_c$ lifetimes. Considering the rough predicted value for the lifetime of $\Xi_{cc}$, one can specify that ${\cal BR}(\Xi_{cc}^+\to \Sigma_{c}^{++}(2520) K^-)\in \left[0.36\%,1.80\%\right]$.

\section*{Acknowledgement}
We are grateful for Ji-Bo He for enlightening discussions which initiated this project. We thank Yu-Ming Wang and Dan Zhang for fruitful discussions. This work was supported in part by the National Natural Science Foundation of China under the Grant Nos. 11505098, 1164700375, 11347027, 11505083, 11575110, 11375208, 11521505, 11621131001, 11235005, 11655002, and 11447032, Natural  Science Foundation of Shanghai under Grant  No. 15DZ2272100 and No. 15ZR1423100,  by the Young Thousand Talents Plan, and Key Laboratory for Particle Physics, Astrophysics and Cosmology, Ministry of Education, and Shanghai Key Laboratory for Particle Physics and Cosmology(SKLPPC), and the Natural Science Foundation of Shandong province ZR2014AQ013, and the Doctoral Scientific Research Foundation of Inner Mongolia University.


\end{document}